\documentclass[]{aastex61}

\received{May 2 2017}
\revised{June 9 2017}
\accepted{June 19 2017}

\begin{document}


\title{OSSOS VI. Striking Biases in the detection of large semimajor axis Trans-Neptunian Objects}

%% information.  If done correctly the peer review system will be able to
%% automatically put the author and affiliation information from the manuscript
%% and save the corresponding author the trouble of entering it by hand.
%%
%% The \affil should be used to document primary affiliations and the
%% \altaffil should be used for secondary affiliations, titles, or email.

%% Authors with the same affiliation can be grouped in a single
\author[0000-0002-3507-5964]{Cory Shankman}
\affiliation{Department of Physics and Astronomy, University of Victoria, Elliott Building, 3800 Finnerty Rd, Victoria, British Columbia V8P 5C2, Canada}
\email{cshankm@uvic.ca}

\author[0000-0001-7032-5255]{J. J. Kavelaars}
\affiliation{NRC-Herzberg Astronomy and Astrophysics, National Research Council of Canada, 5071 West Saanich Rd, Victoria, British Columbia V9E 2E7, Canada}
\affiliation{Department of Physics and Astronomy, University of Victoria, Elliott Building, 3800 Finnerty Rd, Victoria, British Columbia V8P 5C2, Canada}

\author[0000-0003-3257-4490]{Michele T. Bannister}
\affiliation{Astrophysics Research Centre, Queen’s University Belfast, Belfast BT7 1NN, United Kingdom}

\author{Brett J. Gladman}
\affiliation{Department of Physics and Astronomy, The University of British Columbia, Vancouver, BC, V6T 1Z1 Canada}

\author[0000-0001-5368-386X]{Samantha M. Lawler}
\affiliation{NRC-Herzberg Astronomy and Astrophysics, National Research Council of Canada, 5071 West Saanich Rd, Victoria, British Columbia V9E 2E7, Canada}

\author[0000-0001-7244-6069]{Ying-Tung Chen}
\affiliation{Institute of Astronomy and Astrophysics, Academia Sinica; 11F of AS/NTU Astronomy-Mathematics Building, Nr. 1 Roosevelt Rd., Sec. 4, Taipei 10617, Taiwan, R.O.C.}

\author[0000-0002-4385-1169]{Marian Jakubik}
\affiliation{Astronomical Institute, Slovak Academy of Science, 05960 Tatranska Lomnica, Slovakia}

\author{Nathan Kaib}
\affiliation{HL Dodge Department of Physics \& Astronomy, University of Oklahoma, Norman, OK 73019, United States}

\author[0000-0003-4143-8589]{Mike Alexandersen}
\affiliation{Institute of Astronomy and Astrophysics, Academia Sinica; 11F of AS/NTU Astronomy-Mathematics Building, Nr. 1 Roosevelt Rd., Sec. 4, Taipei 10617, Taiwan, R.O.C.}

\author{Stephen D. J. Gwyn}
\affiliation{NRC-Herzberg Astronomy and Astrophysics, National Research Council of Canada, 5071 West Saanich Rd, Victoria, British Columbia V9E 2E7, Canada}

\author[0000-0003-0407-2266]{Jean-Marc Petit}
\affiliation{Institut UTINAM UMR6213, CNRS, Univ. Bourgogne Franche-Comt\'e, OSU Theta F25000 Besan\c{c}on, France}

%\and

\author[0000-0001-8736-236X]{Kathryn Volk}
\affiliation{Lunar and Planetary Laboratory, University of Arizona, 1629 E University Blvd, Tucson, AZ 85721, United States}



\begin{abstract}
The accumulating, but small, set of large semi-major axis trans-Neptunian objects (TNOs) shows an apparent clustering in the orientations of their orbits. 
This clustering must either be representative of the intrinsic distribution of these TNOs, or else arise as a result of observation biases and/or statistically expected variations for such a small set of detected objects. 
The clustered TNOs were detected across different and independent surveys, which has led to claims that the detections are therefore free of observational bias. 
This apparent clustering has led to the so-called ``Planet 9" hypothesis that a super-Earth currently resides in the distant solar system and causes this clustering.
The Outer Solar System Origins Survey (OSSOS) is a large program that ran on the Canada-France-Hawaii Telescope from 2013--2017, discovering more than 800 new TNOs. 
One of the primary design goals of OSSOS was the careful determination of observational biases that would manifest within the detected sample.
We demonstrate the striking and non-intuitive biases that exist for the detection of TNOs with large semi-major axes. 
The eight large semi-major axis OSSOS detections are an independent dataset, of comparable size to the conglomerate samples used in previous studies. We conclude that the orbital distribution of the OSSOS sample is consistent with being detected from a uniform underlying angular distribution.

\end{abstract}




\section{Introduction} \label{sec:intro}

Examining the TNOs in the Minor Planet Center (MPC) database, \citet{trujillosheppard14} noted that the then-known TNOs on orbits with semi-major axis, $a$, beyond 150 au and pericenter, $q$, beyond 30 au have arguments of pericenter, $\omega$, clustered around 0$\degr$ \citep{trujillosheppard14}. 
Many surveys are conducted near the ecliptic plane, and this results in a known bias favoring the detection of TNOs that come to pericenter near the ecliptic plane and thus have $\omega$ near 0$\degr$ or 180$\degr$. 
There has been no demonstrated bias that would favor detections of TNOs with $\omega$ near 0$\degr$ versus those at 180$\degr$. 
\citet{batyginbrown16} noted that the MPC TNOs with $a > 250$ au also have clustered longitude of ascending node, $\Omega$, and longitude of pericenter, $\varpi \equiv \omega + \Omega$. 
Absent additional stabilizing mechanisms, gravitational perturbations from Neptune would randomize these orbital angles on relatively short timescales. 
If the observed clustering of orbital angles is reflective of the intrinsic TNO distribution, there must be some dynamical mechanism forcing these orbital angles to be confined to the present day.
This line of reasoning has led some to hypothesize the existence of an as yet unseen giant planet in the distant solar system to explain the apparent orbital angle clustering \citep{trujillosheppard14,batyginbrown16,malhotraetal16}.
The idea that an unseen planet shapes the distant TNO region is not new and has been invoked to explain the formation of high-perihelion TNOs like (148209) 2000 CR$_{105}$ and (90377) Sedna \citep{gladmanetal02,brownetal04,gomes06,soaresgomes13}.

A key premise of the most recent distant planet hypothesis, which has not yet been independently tested, is that the apparent clustering of orbital angles does not result from observing bias. 
It has been argued that the MPC sample is from independent surveys, thus their biases should be uncorrelated and the observed sample distribution should therefore not have strong biases for the detection of $\omega$ and $\Omega$ \citep{batyginbrown16}.
Unfortunately, most of the TNOs in the MPC are from surveys where the discovery circumstances and survey characteristics remain unpublished, making it impossible to fully account for the observing biases in the full MPC sample.

OSSOS provides a completely independent, single-survey, sample of newly discovered large-$a$ TNOs that is comparable in size to those samples used previously.
OSSOS is a large program on the Canada-France Hawaii Telescope that surveyed 170 deg$^2$ over a range of heliocentric longitudes near the ecliptic in 2013--2017.  
The details of the observing strategy and processing can be found in \citet{bannister16}.
The OSSOS discoveries exceed 830 TNOs with exceptionally well-determined orbits; the high-precision OSSOS astrometry allows rapid orbit determination for classification.
All OSSOS discoveries brighter than the survey flux threshold were carefully and thoroughly tracked to avoid ephemeris biases \citep{jones10}. 
The sensitivity (as a function of flux and motion rate) for each OSSOS observation block is accurately determined, allowing detailed modelling of the sensitivity of OSSOS to TNO orbit distributions. 

The OSSOS large-$a$ TNOs (Table~\ref{tab:TNOlist}) were all detected comparatively close to their perihelia, an expected discovery bias for large-$a$ TNOs.  
OSSOS detected 8 TNOs with $a > 150$ au, $q > 30$ au versus the 12 TNOs from unpublished surveys contained in the MPC used in \citet{trujillosheppard14}, and OSSOS detected 4 TNOs with $a > 250$ au, $q > 30$ au versus the sample of 6 MPC TNOs used in \citet{batyginbrown16}.
The OSSOS sample provides an analogue to the MPC sample while, crucially, also providing the detailed characterization necessary to model the observing biases affecting the detection of our discoveries.

This analysis addresses the following questions:
\begin{enumerate}
\item what are the observing biases, particularly those related to the orbital angles $\omega$, $\Omega$, and $\varpi$, in OSSOS for the $a > 150$ au, $q > 30$ au TNO region? 
\item is there evidence in the OSSOS sample, as has been argued for in the MPC sample of TNOs, of clustering in $\omega$ (for $a >150$ au), $\Omega$ ($a >250$ au) or $\varpi$ ($a > 250$ au)? 
\item can we reject the null hypothesis that the intrinsic distributions of $\omega$, $\Omega$, and $\varpi$ are all uniform?

\end{enumerate}

%----------------------Methods-----------------------------

\section{Observations and Methods} \label{sec:methods}

\subsection{OSSOS Observed Sample of large-$a$ TNOs} \label{sec:obs}

To be consistent with \citet{trujillosheppard14} and \citet{batyginbrown16}, we use the following criteria to define our sample of TNOs: $a > 150$ au and $q > 30$ au. 
OSSOS detected 8 TNOs satisfying the above criteria, 4 of which have $a > 250$ au.
The discovery circumstances for two of these TNOs are described elsewhere: o3e39 \citep[2013 GP$_{136}$]{bannister16} and uo3l91 \citep[2013 SY$_{99}$]{bannister17}. The six new TNOs we present here were found during the rest of the survey: all are characterized discoveries with well-quantified detection efficiencies.
The discoveries have $a$ ranging from 150 au to 735 au, and all but one have $q > 37$ au (Table~\ref{tab:TNOlist}).

\floattable 
\begin{deluxetable*}{ccLCCDDDDCCCCCR} % {lc * {12}{c}}
\tabletypesize{\footnotesize}
\tablecaption{The OSSOS sample of TNOs with $a > 150$ au and $q > 30$ au  \label{tab:TNOlist}}
\tablehead{
%object  a           e         0        i       Omega    omega    dist    mag    Hsur  tperi arc
\colhead{MPC} & \colhead{OSSOS} &  \colhead{$a$} &  \colhead{$e$} &  \colhead{$q$} &  \twocolhead{$i$}  &  \twocolhead{$\Omega$} &  \twocolhead{$\omega$} & \twocolhead{$\varpi$} & \colhead{$r$} & \colhead{$m_r$} &  \colhead{$H_r$} & \colhead{T$_{\textrm{peri}}$} & \colhead{No.} & \colhead{Arc} \\
\colhead{Desig.} & \colhead{Desig.} &  \colhead{(au)} & \colhead{}  &  \colhead{(au)} &  \twocolhead{($^{\circ}$)} &  \twocolhead{$(^{\circ})$} &  \twocolhead{$(^{\circ})$} &  \twocolhead{$(^{\circ})$} &  \colhead{(au)} & \twocolhead{(discovery)} &  \colhead{(JD)} & \colhead{obs.} & \colhead{(days)} \\
}
%\colnumbers

\decimals
\startdata
2013 GP$_{136}$  & o3e39  & 150.2 \pm 0.1 & 0.727 & 41.0 & 33.5 & -149.3 & 45.4  & -106.8 & 45.5 & 23.1 & 6.4 & 2465012 & 31 & 1566 \\
2015 KH$_{163}$ & o5m85  & 153.0\pm0.3 & 0.739 & 39.9 & 27.1 & 67.6   & -129.2 & -61.6 & 51.7 & 24.7 & 7.6 & 2471713 & 36 & 1085 \\
2013 UT$_{15}$ & o3l83  & 200 \pm 1   & 0.780 & 43.9 & 10.7 & -168.0 & -107.9 &  84.1 & 61.2 & 24.1 & 6.2 & 2476001 & 38 & 1278\\
2015 RY$_{245}$ & o5s13  & 226 \pm 3   & 0.861 & 31.4 & 6.0  & -18.5  &   -5.5 & -24.0 & 34.3 & 24.6 & 9.1 & 2452363 & 27 & 538 \\
2015 GT$_{50}$ & o5p060 & 312 \pm 2   & 0.877 & 38.4 & 8.8  & 46.1   & 129.0 & 175.1 & 41.0 & 24.5 & 8.3 & 2451593 & 34 & 824 \\
2015 RX$_{245}$ & o5t52  & 430 \pm 20  & 0.894 & 45.5 & 12.1 &  8.6   & 65.2  & 73.8  & 62.4 & 24.1 & 6.1 & 2475606 & 33 & 587 \\
2015 KG$_{163}$ & o5m52  & 680 \pm 2   & 0.940 & 40.5 & 14.0 & -140.9 & 32.1  & -108.8 & 41.1 & 24.3 & 8.1 & 2459752 & 29 & 739 \\
2013 SY$_{99}$ & uo3l91 & 735 \pm 15 & 0.932 & 50.0 & 4.2  & 29.5  & 32.2  & 61.7 & 60.9 & 24.8 & 6.8 & 2471634 & 33 &  1156 \\
\enddata
\tablecomments{Ordered by semimajor axis, we provide barycentric J2000 ecliptic orbital elements, from the best fit using the method of \citet{bernstein00} to CFHT astrometry listed at the Minor Planet Center as of the time of
publication.
The observed $r$-band magnitude $m_r$, absolute magnitude $H_r$, time of pericenter passage T$_{\textrm{peri}}$, number of observations and length of observed arc are given, along with the barycenter distance $r$ at discovery.
The $1\sigma$ uncertainty from the orbital fit's covariance matrix are listed for $a$; precisions are 0.001 for $e$ and to $0.1\degr$ for the angular elements. All digits presented are significant.}
\end{deluxetable*}

\subsection{A Note On $q$ Selection Criteria}

One might be tempted to impose a $q$ cut higher than 30 au on the sample to select only TNOs that do not have strong gravitational interactions with Neptune (found to be those with $q \la$ 37 au by \citet{lykawkamukai07}).
The argument being that TNOs with $q$ sufficiently close to Neptune will undergo evolution in $a$ and orbital orientation angles on short timescales and so should be removed from the sample.
Embedded in this argument is the assumption that the observed clustering for the sample described above does not result from observation bias.
This work seeks to test that assumption, and so we employ a $q$ lower limit of 30 au to be consistent with prior studies and the region where the MPC TNOs show clustering in orbital angles.

It has also been suggested that only TNOs presently dynamically stable with respect to Neptune should be used to define the sample that is examined for clustering \citep{batyginbrown16}.
The idea being that the TNOs under consideration should be stable to perturbations from Neptune if one is to invoke an additional planet to explain their apparent clustering.
If there is a massive planet in the distant solar system, the region between this planet and Neptune would be, in general, unstable. 
Such a planet would cause pericenter cycling and give dynamical kicks to these large-$a$ TNOs, creating a population analogous to the centaurs, which is seen in multiple simulations with a variety of additional planet candidates \citep{batyginbrown16b,lawleretal16,shankman17}.
TNOs beyond Neptune that are ``presently'' stable would not necessarily be stable in the case of an additional massive planet beyond Neptune.

In any case, if the $q$ threshold is set to a higher limit, one {\it must} still be able to explain why TNOs with $q$ further in, that should be less stable, appear clustered in the MPC sample. 
To reiterate, this analysis examines observational biases and looks for evidence of clustering in the OSSOS sample. 
Thus we select our sample using the same orbital element ranges for which arguments of clustering have been made.

\subsection{Survey Simulation of the Observability of Large-$a$ TNOs}
\label{sec:ssim}

OSSOS is a characterized survey with measured and reported biases.
The pointing directions of the survey itself (Table~1, \citet{bannister16}) are of key importance for the observing biases in orbital angles, as we will demonstrate.
For the purposes of this analysis, we provide the relevant OSSOS TNOs (Table~\ref{tab:TNOlist}) and a full implementation of the survey simulator including an example model distribution is available by request.

We perform simulated surveys on a set of distributions of orbits with $a > 150$ au, $q > 30$ au to probe the effects of the OSSOS observing biases on the detectability of TNOs in the phase space of interest. 
A detailed description of this established survey simulation suite can be found in \citet{jones06} and \citet{petit11}, and recent examples of the use of the survey simulator in TNO studies can be found here: \citet{nesvorny15,alexandersen16,shankman16,pike17}.

We construct test distributions that fully cover ranges of orbital phase space that include the detected large-$a$ TNOs.
The models tested are not intended to reproduce the observed distributions. 
They were designed to probe a variety of forms of distributions to test the sensitivity of the analysis to the specific choice of distribution.
The models tested are combinations of distributions covering the following parameter spaces and forms:
\begin{itemize}
\item $a$: distributions spanned 150 au to 1000 au. Distributions were either uniform in $a$ or $\propto a^{x}$, with exponents $x$ spanning 0.5 to 1.
Distributions with an upper limit of 800 au were also tried to test for sensitivity to the $a$ cut off.
\item eccentricity, $e$: uniform from 0.7 to 0.95. 
A $q$ lower limit was imposed at 30 au.
\item inclination, $i$: Two forms were tested. \textbf{1} a uniform distribution from 0$\degr$ that extends up to 55$\degr$ (the range of the observed OSSOS sample) and \textbf{2} a distribution that scales as $\textrm{sin}(i)\times$gaussian \citep[as in][]{brown01}. 
A variety of gaussian centers (between 0$\degr$, and 20$\degr$) and gaussian widths (between 5$\degr$ and 15$\degr$) including different combinations of centres and widths were used.  
\item absolute magnitude, $H$: single slope from $H_r$ of 6 to 9.5 with a slope of 0.9. Divot and knee distributions as in \citet{fraser14} and \citet{shankman16} were also tested.
\item $\omega$, and $\Omega$ uniform from 0$\degr$ to 360$\degr$, making $\varpi$ uniform as well.
\end{itemize}

With each distribution, we conducted an OSSOS survey simulation that ``detected" 10 000 simulated TNOs. 
These survey simulations reveal the observing biases present in the survey and show any gaps or preferences in the sensitivity to certain orbits.
We find that the choice of model does not affect the conclusions about the intrinsic orbit angle distribution (see Appendix Figure~\ref{fig:mult_models}).

\section{Results} \label{sec:results}

\subsection{Observing Bias} \label{sec:obsbias}

\begin{figure*}[h]
\centering
\includegraphics[width=\textwidth]{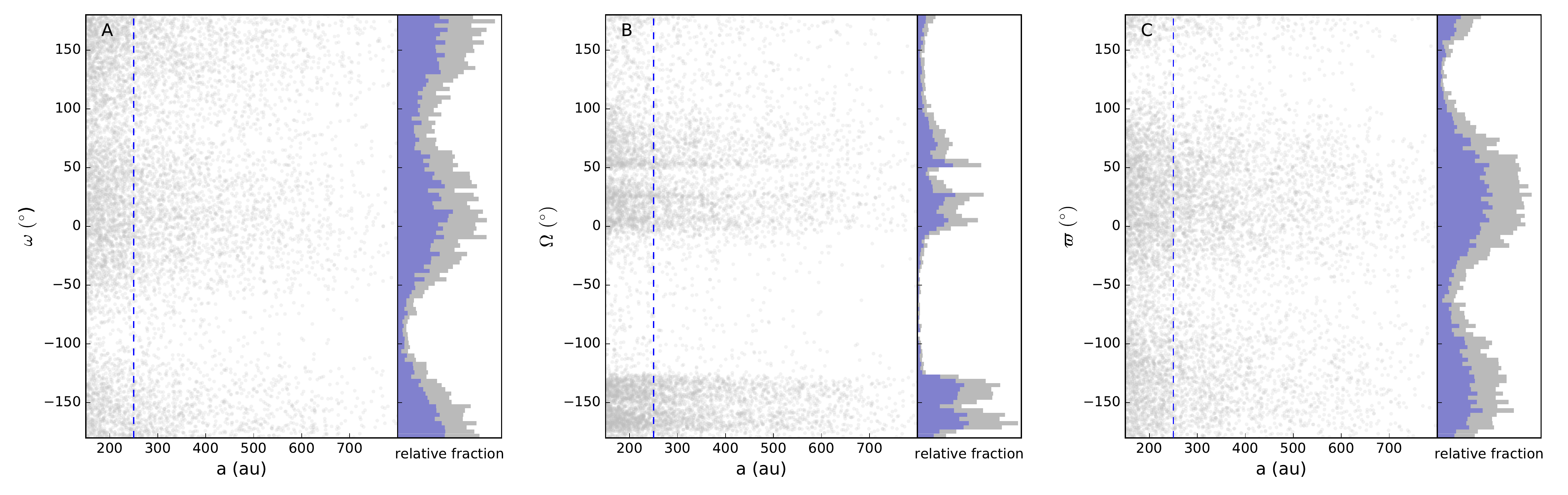}
\caption{Visualization of the detection bias for large-$a$ TNOs in OSSOS.
Simulated detections drawn from a uniform intrinsic distribution are plotted in transparent grey points.
Blue dashed lines in Panels \textbf{A}--\textbf{C} demarcate $a = 250$ au.
The side panels show histograms of the $\omega$, $\Omega$, and $\varpi$ of the orbits of simulated detections.
The grey histograms show the simulated detections of a model with uniform orbit angles. 
Blue histograms show the subset of those simulated detections with $a > 250$ au orbits.
Appendix Figure~\ref{fig:q_hists} shows that the biases do not vary as a function of $q$.
}
\label{fig:bias}
\end{figure*}

Figure~\ref{fig:bias} plots the results of the survey simulation. 
Our simulations find OSSOS has a range of sensitivities to and biases in different orbital parameters of TNOs. 
We discuss in turn our sensitivity to each angle of TNO orbit orientation.
All discussions of panels in this section refer to panels in Figure~\ref{fig:bias}.
Panels \textbf{A}, \textbf{B}, and \textbf{C} plot the orbital angles versus $a$ with histograms of the simulated detections for these angles. 
All statements of the sensitivity in OSSOS are made exclusively with respect to the TNO model constraints as outlined above.

\textbf{$\omega$ sensitivity:}
OSSOS has some sensitivity to all argument of pericenter $\omega$ values, which can be seen in panels \textbf{A}.
Panel \textbf{A} shows that the TNOs on orbits with $\omega$ values near $0\degr$ or 180$\degr$ are more likely to be detected.
This effect arises in near-ecliptic surveys when TNOs are detected near their pericenter.
The OSSOS pointings were not centered exactly around the ecliptic, with almost all the off-ecliptic coverage being North of the ecliptic.
Blocks that are off ecliptic no longer have symmetric sensitivity with $0\degr$ and $180\degr$ favored, but instead have only one area of reduced sensitivity, as also discussed in \citet{sheppardtrujillo16}.
This lack of sensitivity is caused by the fact that some pericenter locations are not possible to detect for a survey off ecliptic.
For example, a survey that points above the ecliptic is unable to see any sky points below the ecliptic and thus has a lower sensitivity to detecting orbits that come to pericenter below the ecliptic.
This effect results in less OSSOS sensitivity to TNOs with $\omega$ near $-90\degr$ than near $90\degr$. 
OSSOS still has some sensitivity to TNOs away from their pericenter due to its deep limiting magnitudes, resulting in some sensitivity to TNOs with $\omega$ near $-90\degr$, as seen in panel \textbf{A}.

\begin{figure*}[h]
\centering
\includegraphics[width=0.5\textwidth]{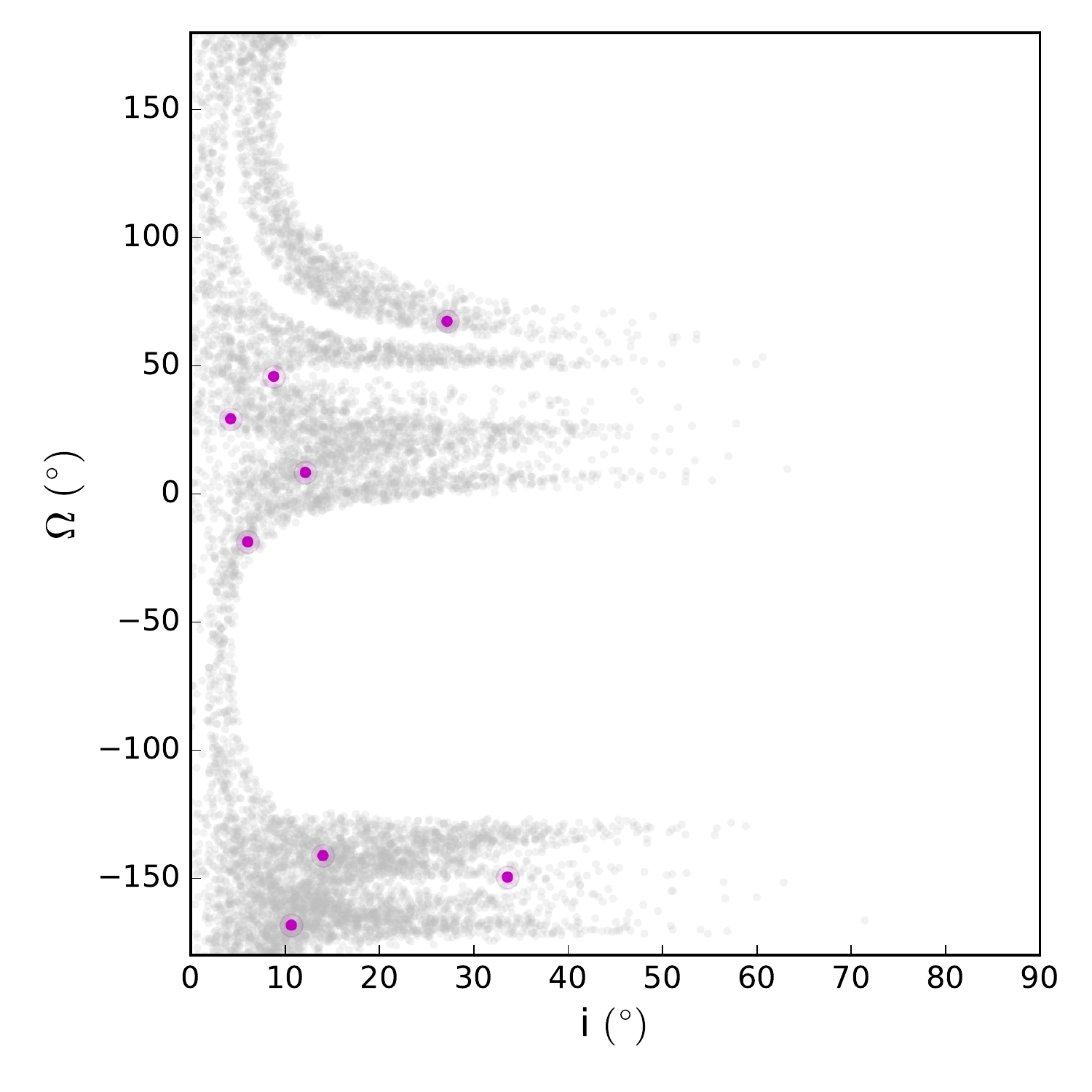}
\caption{
The $\Omega$ / $i$ sensitivity of the OSSOS project.  
The grey dots are the same simulated detections in Figure~\ref{fig:bias}, having $a > 150$ au, $q >$ 30 au.
The OSSOS TNOs have been overplotted in magenta to show how the observed sample is affected by these strong biases.
}
\label{fig:Omega_bias}
\end{figure*}

\textbf{$\Omega$ sensitivity:}
There exists clear and initially non-intuitive structure in the OSSOS observing bias in longitude of the ascending node $\Omega$.
Panel \textbf{B} shows that there is a large and substantial gap in $\Omega$ sensitivity in the -120$\degr$ to -20$\degr$ range.
OSSOS, due to its avoidance of the galactic plane and northern hemisphere winter, has virtually no capability for detecting large-$a$ TNOs with $\Omega$ between -120$\degr$ to -20$\degr$. 
Figure~\ref{fig:Omega_bias} shows that this structure arises from a coupling of sensitivity in $\Omega$ and $i$.
This striking effect is a simple result of geometry.
The $\Omega$ and $i$ angles define the plane of the TNO's orbit. 
In order for the TNO to be detectable by a survey, its plane must intersect the area of sky being observed. 
For inclined orbits to have a high chance of being detectable in an ecliptic survey, the ascending or descending node must be in the same direction as the survey's pointing. 
Each of the horizontal spikes in Figure~\ref{fig:Omega_bias} indicates the location of an ascending or descending node that is aligned with one of the OSSOS pointings. 
The bias structure curves horizontally as inclinations go towards 0$\degr$, when orbits become ecliptic grazing, and can thus be seen at more points across their orbit in ecliptic surveys. 
We plot the OSSOS detections in Table~\ref{tab:TNOlist} in panel Figure~\ref{fig:Omega_bias} to show that they follow this bias-induced distribution of $\Omega$ and $i$.

\textbf{$\varpi$ sensitivity:}
The sensitivity to detecting longitude of pericentre $\varpi$ is a combination of the sensitivities to detecting $\Omega$ and $\omega$.
The sensitivity is a double peaked distribution, with less sensitivity to $\varpi$ in the range of 110$\degr$ to 160$\degr$ (see Panel \textbf{C}).
OSSOS has some sensitivity to all $\varpi$ values and there is no striking structure in the $\varpi$ sensitivity other than the two broad peaks roughly separated by 180$\degr$ of longitude.

We examined the three above biases for both the $a > 150$ au and $a > 250$ au regions to explore if the sensitivity changes with $a$. 
We find that the observing biases are the same for the two regions, which can be seen by comparison of the blue and grey histograms in Figure~\ref{fig:bias}.
Once a TNO has a sufficiently large orbit, it is only detectable near pericenter. 
This bias strongly affects the expected detection of orbital angles as we have shown.
The bias structure does not change as a function of increasing $a$ because the bias is a result of the fact that the TNOs are only detectable near pericenter.

Although the biases we demonstrate are specific to OSSOS, all TNO surveys will have complicated detection biases like those shown in this work.  
Without publishing characterizations of the survey pointings, these complex and often non-intuitive biases cannot be accounted for, and may lead to incorrect assumptions about the intrinsic population.

%-----Figure------

%-----Figure------

%-----Figure------

%-----Figure------

\subsection{Angle Clustering in the OSSOS Sample}

Having examined the biases of OSSOS, we now examine the detected OSSOS sample (Table~\ref{tab:TNOlist}) for evidence of a clustering in the orbital angles for the large-$a$ TNOs, as first noted in the MPC dataset by \citet{trujillosheppard14} and \citet{brownbatygin16}.
We consider the OSSOS sample independently, examining the distributions of the OSSOS TNOs with no \textit{a priori} expectations about clustering.

\begin{figure*}[h]
\centering
\includegraphics[width=\textwidth]{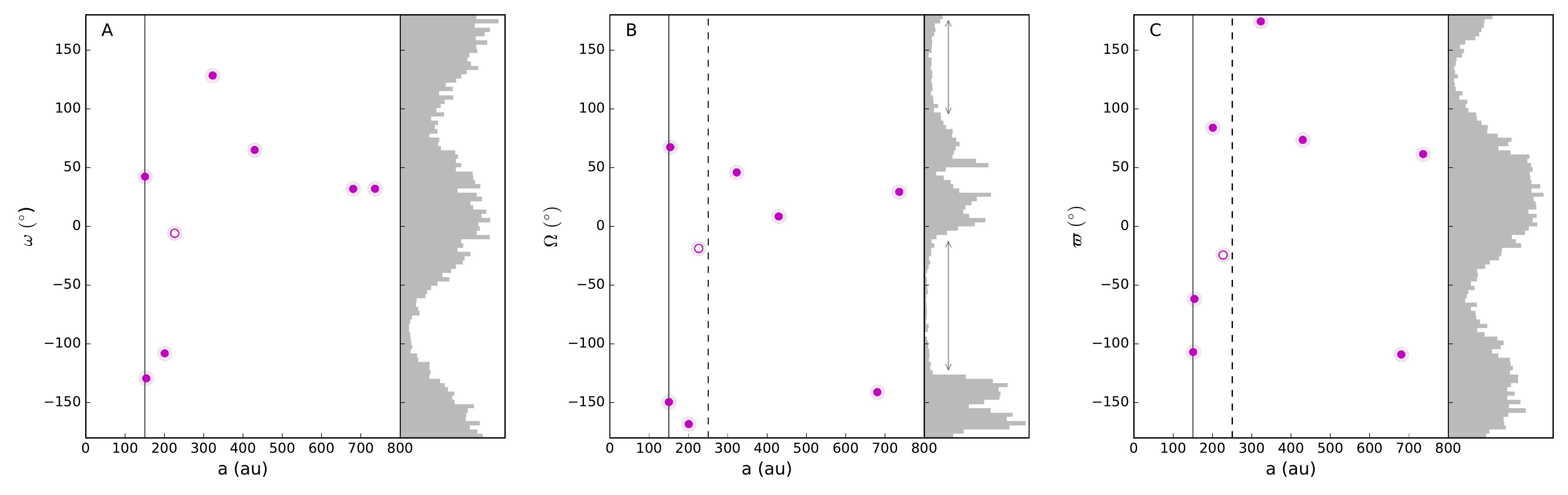}
\caption{
The OSSOS detections satisfying $a > 150$ au and $q > 30$ au are shown with magenta points.
Open circles indicate TNOs with $q < 37$ au - a threshold cited to demarcate the region of stability from Neptune perturbations \citep{lykawkamukai07}.
Solid vertical lines mark 150 au and dashed lines mark 250 au.
Histograms repeat the OSSOS sensitivity in each parameter as in Figure~\ref{fig:bias}.
Double sided arrows in the Panel B histogram mark the $\Omega$ ranges where OSSOS has low sensitivity due to the survey's bias, and thus detections are unlikely.
}
\label{fig:angles_OSSOS}
\end{figure*}

A visual examination of the orbital distribution (see Figure~\ref{fig:angles_OSSOS}) shows relatively little evidence of clustering of $\omega$, even in the observationally biased $a > 150$ au OSSOS sample.
The eight OSSOS TNOs are found distributed across the full range of $\omega$ values (Figure~\ref{fig:angles_OSSOS} panel \textbf{A}).
We demonstrated in Section~\ref{sec:obsbias} that OSSOS has some sensitivity to all $\omega$ values; this sensitivity is reflected in the broad distribution of detected $\omega$ values.
For each orbital angle, we test the hypothesis that the OSSOS sample can be detected from a uniform intrinsic distribution.
To do this, we compare the observed distributions to the survey simulator biased models described above (Figure~\ref{fig:bias}).
We test the null hypothesis using Kuiper's test \citep[e.g. see][]{fisher95,pewsey13}, which is closely related to the Kolmogorov-Smirnov test, but is invariant to cyclic transformations of the test variable.
The test is thus well suited to problems with cyclic variables, as is the case for angles like $\omega$, $\Omega$, and $\varpi$.
The eight TNOs are statistically consistent (i.e. 53\% of bootstrapped model subsamples have a larger Kuiper's test distance than the observed sample to the parent model: the hypothesis is rejectable at 47\%, i.e. not rejectable) with being detected from an intrinsic uniform distribution of $\omega$ values.
Our results hold for all tested intrinsic models described in Section~\ref{sec:ssim}.

We now examine the $\Omega$ and $\varpi$ distributions for the OSSOS TNOs with $a > 250$ au.
We find that the $\Omega$ values for three of the TNOs are distributed near 25$\degr$ with the fourth isolated (Figure~\ref{fig:angles_OSSOS} panel \textbf{B}).
OSSOS had effectively no sensitivity to TNOs with $\Omega$ between -120$\degr$ and -20$\degr$, and had poor sensitivity to TNOs with $\Omega$ between 115$\degr$ and 165$\degr$ (see Figure~\ref{fig:angles_OSSOS} panel B histogram).
Unsurprisingly, OSSOS did not detect TNOs in regions of limited or no sensitivity.
Using Kuiper's test, we find that the OSSOS detections are statistically consistent (rejectable at 61\%, i.e. not rejectable) with being detected from an intrinsic uniform distribution of $\Omega$ values.
We find that the $\varpi$ values cover a large range, with only two values near each other.
As with $\omega$ and $\Omega$, the OSSOS TNO $\varpi$ values are consistent (rejectable at 62\%, i.e. not rejectable) with being detected from an intrinsic uniform distribution of $\varpi$ values.

We conclude that the independent OSSOS sample shows no evidence for intrinsic clustering in the $\omega$, $\Omega$ or $\varpi$ distributions of TNOs.

\subsection{OSSOS and MPC Sample Comparison}
\label{sec:OSSOS_MPC}

\begin{figure*}[h]
\centering
\includegraphics[width=1.0\linewidth]{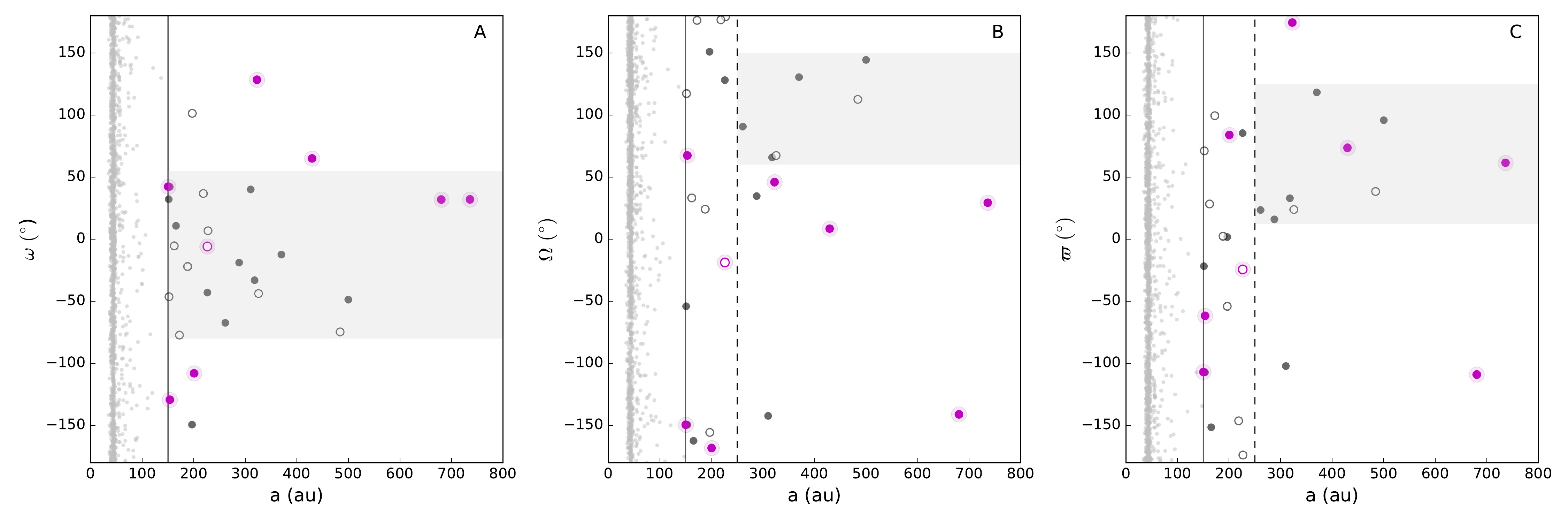}
\caption{
Plots of the orbits of the OSSOS and MPC samples for $\omega$, $\Omega$ and $\varpi$ versus $a$.
The MPC sample\footnote{2014 FE$_{72}$ with $a = 2155$ au, $q = 36$ au, $i = 20\degr$, $\omega = 134\degr$, and $\Omega = -23\degr$ has been excluded from these plots because it interacts with galactic tides \citep{sheppardtrujillo16}} has been selected with $q > 30$ au.
OSSOS detections are shown in magenta points.
MPC TNOs are plotted in transparent grey points, which are larger for $a > 150$ TNOs.
There are no OSSOS discoveries in the $a > 150$ au MPC grey points.
Solid vertical lines mark 150 au and dashed lines mark 250 au.
The grey shaded regions indicate the regions of apparent clustering in the MPC sample proposed by previous authors.
As in Figure~\ref{fig:angles_OSSOS}, open circles indicate TNOs with $q < 37$ au, showing how a stability argument might affect the argument for clustering. 
It is clear from this view that the MPC TNOs with $q$ between 30 au and 37 au still appear to cluster in $\omega$ (for $a > 150$ au) and $\Omega$ and $\varpi$ (for $a > 250$ au), despite the fact that interactions with Neptune would prevent shepherding by an external planet.
}
\label{fig:angles_MPC}
\end{figure*}

We now compare the OSSOS sample (known biases) to the MPC sample (unknown biases) to examine the broader question of clustering in the known TNOs.
Figure~\ref{fig:angles_MPC} plots the eight OSSOS TNOs and the MPC TNOs satisfying $a > 150$ au and $q > 30$ au as reported by the MPC in April 2017.

Figure~\ref{fig:angles_MPC} panel \textbf{A} shows clearly that the apparent clustering in $\omega$ that has been noted in the MPC sample is not present in the OSSOS sample, despite the OSSOS survey biases against $\omega = \pm 90\degr$. 
Where the MPC sample is contained within roughly 50$\degr$ of 0$\degr$, the OSSOS sample spans all values and has as many TNOs inside the apparent clustering region (grey shading of Figure~\ref{fig:angles_MPC} panel \textbf{A}) as outside.
The OSSOS TNOs that are outside the apparent clustering region have a variety of semi-major axes and all have $q > 37$ au.
With the addition of the OSSOS sample and a few recently discovered TNOs in the MPC sample, the argument for a clustering of $\omega$ in the detected TNOs has been substantially weakened. 

There is no overlap between the OSSOS sample and the $\Omega$ clustering region of TNOs with $a > 250$ au noted in \citet{batyginbrown16}, with all four OSSOS detections outside the clustered band.
The four $a > 250$ au OSSOS detections span a range of $a$ values and all have $q > 37$ au.
If one were to consider the three OSSOS detections with $\Omega$ between 0$\degr$ and 50$\degr$ to be part of the clustered grouping, the clustering would then span approximately 150$\degr$.
\citet{sheppardtrujillo16} noted that their recent discoveries began to erode the signal of clustered $\Omega$ in the large-$a$ TNOs; the four OSSOS detections outside the previously reported band continue this trend of eroding the signal. 
This would be expected if the original signal results from a combination of small number statistics and observing bias.

Two of the OSSOS $a > 250$ au TNOs with $\varpi$ near $70\degr$ are within the \citet{batyginbrown16} proposed ``anti-aligned'' cluster.
The third, o5m52 with a $\varpi$ of $-110\degr$ is approximately 180$\degr$ away, and would fall in the subsequently postulated ``aligned'' cluster \citep{brownbatygin16,sheppardtrujillo16}.
Discoveries at these values of $\varpi$ is unsurprising in OSSOS as the observing bias favors detections with $\varpi$ at these longitudes  (Figure~\ref{fig:bias} panel \textbf{C}).
The final OSSOS detection, o5p060, however, is approximately 90$\degr$ away from each of these proposed clustering regions.

%-----Figure------

%-----Figure------

%-----Figure------

%-----Figure------

%-----Figure------

\begin{figure*}[h]
\centering
\includegraphics[width=0.9\textwidth]{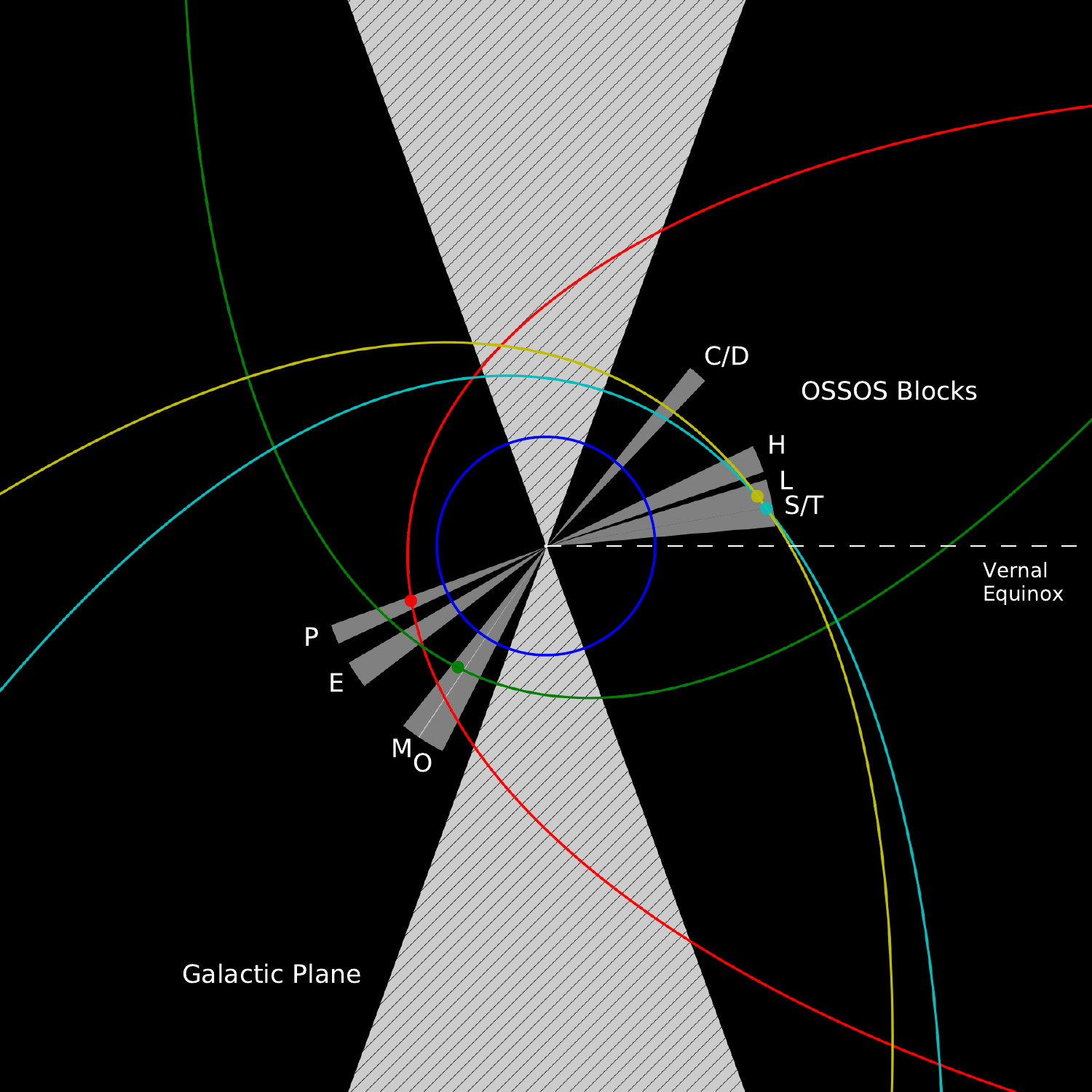}
\caption{A top-down view of the Solar System including Neptune, a schematic for the OSSOS pointings and the four $a > 250$ au OSSOS TNOs.
Neptune's orbit is plotted with a blue circle.
The OSSOS TNOS are plotted in the following colours:
o5p060 red,
o5m52 green,
o5t52 cyan,
uo3l91 yellow.
The discovery location of each TNO is indicated by a point of the appropriate colour.
The eight OSSOS blocks \citep[][Table 1]{bannister16} are plotted in grey and labelled (note that detection sensitivity continues radially beyond the wedge boundaries).
The rough location of the galactic plane is plotted in hatched wedges.
A dashed line indicates the direction of the vernal equinox, and therefore the upper right quadrant is the September to November opposition direction.
}
\label{fig:topdown}
\end{figure*}

%-----Figure------

\section{Discussion and Summary}
\label{sec:discussion}

We find no evidence in the OSSOS sample for the $\omega$ clustering that was the impetus for the current additional planet hypothesis \citep{trujillosheppard14}.
The OSSOS $\omega$ distribution cannot reject the null hypothesis that the underlying distribution is random, once the biases are taken into account.
Our analysis of the OSSOS survey bias and our detections do not directly address the question of why the majority of presently known MPC TNOs are clustered around $\omega$ of 0$\degr$. 
We suggest that this clustering is the result of a combination of observing bias and small number statistics, though we cannot test this without published characterizations of the surveys that detected these TNOs. 
It must be the case that OSSOS and the other surveys that compose the MPC sample have observed the same intrinsic distribution.
OSSOS found TNOs across all values despite being most sensitive to TNOs in the clustering band (near 0$\degr$).
The OSSOS detections go beyond the relatively tight clustering seen in the observed sample, and the OSSOS distribution is consistent with a uniform intrinsic $\omega$ distribution.
This result calls into question the idea of a clustering of $\omega$ around $0\degr$ in the intrinsic distribution of $a > 150$ au $q > 30$ au TNOs.

We have demonstrated that $\Omega$ biases are strong and very present in surveys such as OSSOS. 
These complex biases must also exist in the surveys that compose the MPC sample; it is not sufficient to state that the surveys are independent and therefore the biases must have averaged out.
There have been only a handful of surveys that have detected such large-$a$ TNOs, and the biases from these surveys have shaped the MPC sample in unknown ways. 
There is a large gap in the known TNO $\Omega$ distribution for both OSSOS and MPC samples.
This gap occurs precisely where OSSOS has no sensitivity due to the survey's construction. 
In OSSOS, this gap is driven by weather patterns at the Canada-France-Hawaii Telescope and pointing choices that avoid the dense star fields of the galactic plane. 
Figure~\ref{fig:topdown} provides a visual representation that demonstrates the nature of these biases.
OSSOS observed in the northern spring (April--May) and fall (September--November), and has virtually no sensitivity to orbits that intersect the ecliptic at other times of the year.
TNO surveys have been conducted from a limited number of locations and are subject to similar constraints as OSSOS.
In particular, the best conditions (and thus deepest coverage) is in these months.
It is therefore possible that the surveys which detected the MPC sample contain these same biases and therefore the gap in the detected $\Omega$ distribution may result simply from pointing constraints.

It has been argued that no $\omega$ or $\varpi$ biases are seen in the close-in TNOs in the MPC sample and thus there should be no biases in the large-$a$ sample.
Unique biases arise from the fact that the large-$a$ TNOs are only detected near their pericenters. 
The lack of observed clustering in the close-in TNOs cannot simply be extended to conclude that the large-$a$ TNOs lack bias.
To verify this via simulation, we examined the OSSOS sensitivity to close-in TNOs, as in \S~\ref{sec:ssim}. 
OSSOS has equal sensitivity for all values of $\omega$ and $\varpi$, in contrast to the striking biases observed for the large-$a$ TNOs (see Appendix Figure~\ref{fig:close_model}).

Much attention has been given to the appearance of clustering of ``aligned'' and ``anti-aligned'' orbits in physical space (apparent clustering of $\varpi$), which is founded on the assertion that there is no bias in the detection of $\varpi$.
However, we have shown that for TNOs detected near pericenter, the detection of $\varpi$ is also driven strongly by where one looks in the sky.
Historical surveys will also show bias favoring detections in the aligned and anti-aligned directions due to the prevalence of large-scale surveys occurring (and having the best weather) in the spring and fall.
In order to efficiently detect large-$a$ TNOs outside the aligned and anti-aligned clusters, one would need large survey coverage during June--August and December--February. 
We posit that two clusters 180$\degr$ apart are the natural outcome of seasonal weather biases when observing a highly eccentric population for which detection is only possible close to pericenter.
The observed TNOs therefore do not require the existence of a non-uniform intrinsic distribution (the impetus of the additional planet hypothesis).
Additionally, the MPC sample's $\omega$ values are all near 0$\degr$ and as we have shown, the detected $\Omega$ distribution is strongly set by the geometry of the pointing directions. 
The clustering seen in $\varpi$ in the MPC sample, therefore likely results from adding numbers near 0$\degr$ to strong observing biases present in the $\Omega$ distribution.
The apparent $\varpi$ clustering seen in the MPC sample thus cannot be taken to be independent of bias.

While OSSOS was primarily sensitive to orbits with $\varpi$ near the region of the MPC sample clustering (see Figure~\ref{fig:bias}), it still found one quarter of its sample away from this region where the sensitivity is low.
Despite the reduced sensitivity to such orbits, OSSOS detected o5p060 with a $\varpi$ that produces an orbit orthogonal to the suggested clustering axis. 
This suggest that there must be a large population of TNOs on similar orbits (of order ten thousand\footnote{Survey simulations show that approximately 13,000 TNOs with $H_r < 9$ on orbits within the uncertainty of o5p060's orbit are required to explain the detection of o5p60 in OSSOS.}), or that the detection of o5p060 was anomalous. 
In either case, the existence of o5p060 with $a$ of 314 au, $q$ of 38 au, and $\varpi$ 90$\degr$ away from the clustering region provides evidence of a population that would refute a simplistic interpretation of the extra-planet hypothesis \citep{brownbatygin16} in which \emph{only} anti-aligned orbits can survive.

One might be tempted to choose a different pericenter sample cut, pushing the limit away from 30 au. 
Setting a limit of 40 au would remove two of the six TNOs noted by \citet{batyginbrown16} to cluster and one of the four in the OSSOS sample. 
If one is to argue that a dynamical effect causes the clustering of only the TNOs with $q$ greater than a limit of 40 au (or any other choice), {\it it must then be explained} why the MPC sample of TNOs with $ 40 < q < 30$ au also appear to cluster, if the effect is not caused by observing bias.

We have shown that there are strong and striking biases in the detection of the orbital angles present in OSSOS.
There is no evidence for clustering in the OSSOS sample when considered alone, and when OSSOS is folded into the MPC sample the arguments for clustering in the detected TNOs erodes. 
The first large independent sample shows no evidence for the hypothesized intrinsic clustering.
While the idea of there being a larger-than-dwarf-scale planet in the outer solar system as a mechanism to create the $q$-detached TNOs is still plausible, the evidence that there is currently a super-Earth or larger planet confining the large-$a$ TNOs is in doubt.

\acknowledgments

This project was funded by the National Science and Engineering Research Council and the National Research Council of Canada. This research used the facilities of the Canadian Astronomy Data Centre operated by the National Research Council of Canada with the support of the Canadian Space Agency.
CJS gratefully acknowledges support from the NSERC CGS Fellowship.
MTB appreciatively acknowledges support from UK STFC grant ST/L000709/1.
SML gratefully acknowledges support from the NRC Canada Plaskett Fellowship.
M.J. acknowledges the support from the Slovak Grant Agency for Science (grant No. 2/0031/14).
The authors recognize and acknowledge the sacred nature of Maunakea, and appreciate the opportunity to use data observed from the mountain. 
We thank the dedicated staff of the Canada--France--Hawaii Telescope (CFHT). 
CFHT is operated by the National Research Council of Canada, the Institute National des Sciences de l'Univers of the Centre National de la Recherche Scientifique of France, and the University of Hawaii, 
with OSSOS receiving additional access due to contributions from the Institute of Astronomy and Astrophysics, Academia Sinica, Taiwan.
Observations were obtained with MegaPrime/MegaCam, a joint project of CFHT and CEA/DAPNIA.
We would like to thank the anonymous referee for their discussion and suggestions which have helped us to improve this work.


%% for the paper.  Each keyword is check against the master list during
%% copy editing.  Individual instruments can be provided in parentheses,
%% after the keyword, but they are not verified.

\vspace{5mm}
\facilities{CANFAR, NRC, CFHT (MegaPrime)}

\software{Mercury, Python}


\newpage

\appendix

\section{Additional Figures}

%-----Figure------

\begin{figure*}[h]
\centering
\includegraphics[width=\textwidth]{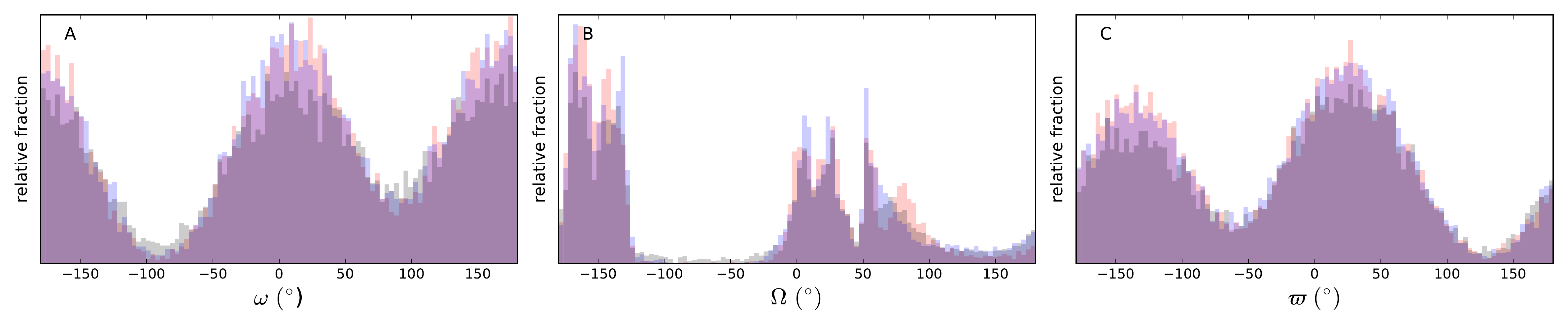}
\caption{Histograms of the sensitivity of OSSOS, as in Figure~\ref{fig:bias}, to three different models of orbit distributions (plotted with transparencies).
These three models explore different $a$ and $i$ distributions.
One model has flat $a$ and $i$ distributions with $a$ spanning 150 au - 800 au and $i$ up to 55$\degr$.
The other two models have power-law $a$ distributions and $i$ distributions that are drawn from $\textrm{sin}(i)\times$gaussian distributions with different centres and widths.
This shows that the results hold in general across different choices of model distributions.}
\label{fig:mult_models}
\end{figure*}

%-----Figure------

%-----Figure------

\begin{figure*}[h]
\centering
\includegraphics[width=\textwidth]{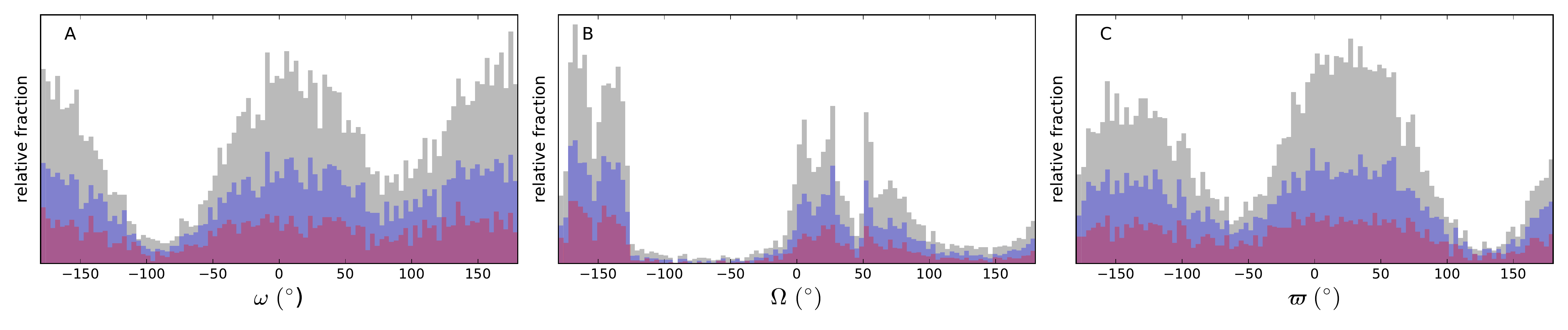}
\caption{Histograms of the sensitivity of OSSOS as in Figure~\ref{fig:bias}.
The grey histogram shows all simulated detections, the blue histogram shows those with $q < 37$ au, and the magenta histogram shows TNOs with an even lower cutoff of $q < 34$ au.
The form of the bias is the same for all $q$ cuts, but the bias is more pronounced for the largest $q$ TNOs.}
\label{fig:q_hists}
\end{figure*}

%-----Figure------

%-----Figure------

\begin{figure*}[h]
\centering
\includegraphics[width=\textwidth]{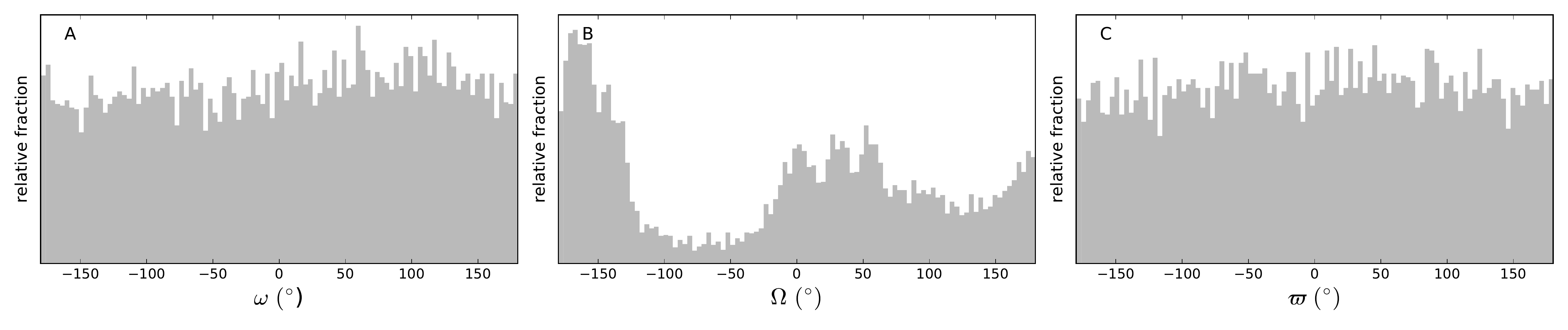}
\caption{Plots of the OSSOS sensitivity to a model orbit distribution with close-in TNOs. 
The model has both $a$ and $q$ between 30 and 50 au.
The strong biases observed in $\omega$ and $\varpi$ for the large-$a$ TNOs (Figure~\ref{fig:Omega_bias}) are not present for this close-in population, which has near equal sensitivity to detecting TNOs with all $\omega$ and $\varpi$ values.
The biases seen in Figure~\ref{fig:bias} arise from detecting TNOs near pericenter, and thus are not present for close-in TNOs which can be detected at any point in their orbit.
The biases in $\Omega$ arise from the geometry of orbits intersecting pointing locations, and so are still present in the close-in sample.}
\label{fig:close_model}
\end{figure*}

%-----Figure------




\bibliographystyle{apj}
\bibliography{citations}


\listofchanges

\end{document}